\documentclass[conference]{IEEEtran}
\IEEEoverridecommandlockouts
\usepackage{cite}
\usepackage{amsmath,amssymb,amsfonts}
\usepackage{algorithm}
\usepackage{algpseudocode}
\usepackage{graphicx}
\usepackage{textcomp}
\usepackage{xcolor}
\usepackage{url}
\def\BibTeX{{\rm B\kern-.05em{\sc i\kern-.025em b}\kern-.08em
    T\kern-.1667em\lower.7ex\hbox{E}\kern-.125emX}}
\begin{document}

\title{Resilient Time-Sensitive Networking for Industrial IoT: Configuration and Fault-Tolerance Evaluation
\thanks{This publication has emanated from research conducted with the financial
support of Research Ireland under Grant number 13/RC/2077 P2.
For the purpose of Open Access, the author has applied a CC-BY public
copyright license to any Author Accepted Manuscript version arising from
this submission}
}

\author{\IEEEauthorblockN{Mohamed Seliem, Dirk Pesch, Utz Roedig, and Cormac Sreenan}
\IEEEauthorblockA{\textit{School of Computer Science and Information Technology} \\
\textit{National Universities of Ireland, University College Cork}, Cork, Ireland \\
MSeliem@ucc.ie, Dirk.pesch@ucc.ie, u.roedig@ucc.ie, Cormac.Sreenan@ucc.ie}
}

\maketitle

\begin{abstract}
Time-Sensitive Networking (TSN) is increasingly adopted in industrial systems to meet strict latency, jitter, and reliability requirements. However, evaluating TSN’s fault tolerance under realistic failure conditions remains challenging. This paper presents IN2C, a modular OMNeT++/INET-based simulation framework that models two synchronized production cells connected to centralized infrastructure. IN2C integrates core TSN features—including time synchronization, traffic shaping, per-stream filtering, and Frame Replication and Elimination for Redundancy (FRER)—alongside XML-driven fault injection for link and node failures. Four fault scenarios are evaluated to compare TSN performance with and without redundancy. Results show that FRER eliminates packet loss and achieves sub-millisecond recovery, though with 2–3× higher link utilization. These findings offer practical guidance for deploying TSN in bandwidth-constrained industrial environments.
\end{abstract}

\begin{IEEEkeywords}
TSN, Industrial Ethernet, OMNeT++, Deterministic Networking, Fault Tolerance, Network Resilience.
\end{IEEEkeywords}

\section{Introduction}
Ensuring deterministic, low-latency, and fault-tolerant communication is vital for industrial automation and control systems. The convergence of Information Technology (IT) and Operational Technology (OT), coupled with the rise of cyber-physical systems, imposes stringent performance and reliability requirements on network infrastructures. While traditional Ethernet remains widespread, it lacks the timing guarantees, traffic isolation, and failure resilience required by real-time industrial applications.

Time-Sensitive Networking (TSN), a suite of IEEE 802.1 standards, augments Ethernet with features such as precise time synchronization, Generalized Precision Time Protocol (gPTP), time aware shaping (TAS), per-stream filtering and policing, VLAN-based prioritization, and Frame Replication and Elimination for Redundancy (FRER). These capabilities collectively enable deterministic behavior, low jitter, and high availability, making TSN a key enabler for industrial Internet of Things (IIoT) and smart manufacturing networks. However, assessing TSN’s effectiveness under real-world failure conditions remains an open research challenge.

This paper presents a modular OMNeT++/INET-based simulation framework that models a realistic, multi-cell industrial network equipped with TSN features. Our simulation framework includes configurable traffic patterns, Virtual Local Area Network (VLAN)-based stream prioritization, gPTP-based synchronization, per-stream filtering, and stream redundancy using FRER. Crucially, the framework integrates XML-based failure scenarios to simulate both link and cell-level disruptions and evaluates TSN's recovery behavior with and without redundancy. The key contributions of this work are:
\begin{itemize}
    \item \textbf{Realistic Industrial TSN Modeling:} A full-scale industrial network with two production cells \cite{b0} is implemented in OMNeT++, integrating TSN mechanisms such as time synchronization, stream filtering, traffic shaping, and redundancy.
    \item \textbf{Scenario-Based Fault Injection:} Link and node failures are modeled using timed XML scripts, enabling the evaluation of FRER and failover mechanisms under controlled disruptions. 
    \item \textbf{Resilience Evaluation:} The framework provides detailed performance insights—including latency, jitter, and recovery time—under both baseline and redundancy-enhanced TSN configurations.
\end{itemize}

The remainder of this paper is structured as follows: Section II reviews related work on TSN for industrial systems. Section III outlines the system architecture. Section IV details the simulation setup, including topology, traffic modeling, and fault injection. Section V presents results across the four scenarios. Section VI concludes and outlines future directions.

\section{Background and Related Work}
TSN has emerged as a key enabler of deterministic Ethernet communication for industrial automation, automotive, and other latency-sensitive domains. Defined by the IEEE 802.1 working group, TSN incorporates core features such as time synchronization (IEEE 802.1AS), time-aware scheduling (IEEE 802.1Qbv), per-stream filtering and policing (IEEE 802.1Qci), and Frame Replication and Elimination for Redundancy (FRER, IEEE 802.1CB) to ensure low-latency, bounded jitter, and high availability of critical traffic flows \cite{b1}. These mechanisms allow Ethernet to meet the stringent requirements of Industrial Internet of Things (IIoT) applications, replacing traditional fieldbus and proprietary real-time solutions \cite{b2}.

Extensive research has confirmed the benefits of TSN in reducing jitter and improving deterministic behavior in mixed-criticality environments \cite{b3}. Simulation-based studies—particularly those using OMNeT++ and the INET framework—have validated the effectiveness of TSN for traffic prioritization, scheduling, and reliability in industrial scenarios \cite{b4}. Fault tolerance, a cornerstone of industrial communication, is typically achieved in TSN using redundant streams and fast fail-over techniques such as FRER \cite{b5}. These features are especially relevant in factory networks where uptime and predictable performance are critical.

While TSN has gained traction in automotive networking \cite{b6}, its deployment in industrial automation—particularly under realistic fault conditions—remains an open research challenge. Recent work has examined TSN’s integration with legacy protocols such as PROFINET and EtherCAT \cite{b7}, and proposed new scheduling strategies for optimizing latency and bandwidth usage \cite{b8}. Beyond scheduling, several studies have focused on fault-tolerance and redundancy in TSN and wireless TSN systems. Sudhakaran et al. \cite{b9} introduced a dynamic scheduling approach enabling zero-delay roaming for mobile robots using wireless TSN redundancy, showing that FRER can uphold real-time guarantees under mobility. Jover et al. \cite{b10} analyzed the trade-off between reliability and redundancy cost in TSN networks through simulation-based configuration insights. On the wireless side, Cena et al. \cite{b11} proposed seamless redundancy techniques for high-reliability Wi-Fi, leveraging Wi-Fi 7 Multi-Link Operation (MLO) to implement FRER-like duplication and de-duplication. Those complement our study by tackling advanced fault-handling approaches. However, most existing efforts focus on static performance or isolated failure events, overlooking complex failure injection scenarios with synchronized industrial traffic. Our work addresses this gap by introducing a scenario-driven simulation framework (IN2C) that models industrial production cells, real-time traffic, and coordinated failure events. It evaluates four distinct failure configurations, each with and without FRER, to quantify TSN’s fault-recovery capabilities in terms of latency, jitter, and recovery time. The novelty of this work lies in:
\begin{itemize}
    \item Comprehensive fault injection across layered industrial traffic flows, using XML-triggered scenarios and real-world-inspired production logic;
    \item Comparative evaluation of TSN with and without redundancy, enabling quantitative insights into the impact of FRER and per-flow protection under failure;
    \item Multi-role traffic modeling (sensor-actuator-PLC-SCADA-Edge unit-HMI), reflecting realistic IIoT interactions rarely combined at this level of detail in existing TSN studies.
\end{itemize}

These contributions provide a deeper understanding of TSN’s resilience in industrial environments and offer a reusable simulation testbed for future TSN extensions and industrial scheduling strategies.

\begin{figure}[t]
    \centering
    \includegraphics[width=\linewidth]{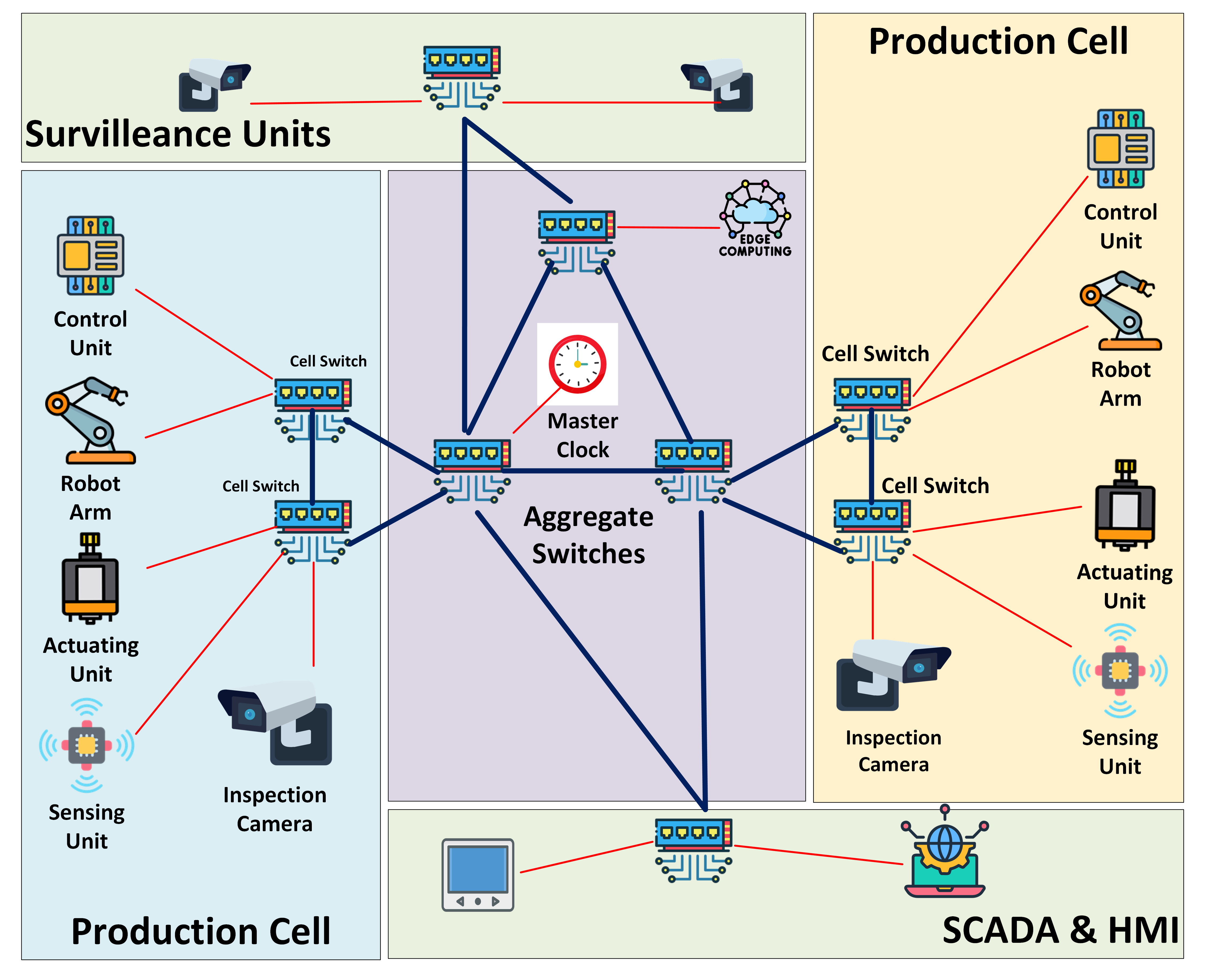}
    \caption{Industrial TSN Network with Two Production Cells}
    \label{fig:network}
\end{figure}

\section{System Model and Design Considerations}
The industrial TSN network designed in this work models a realistic smart factory with two production cells and centralized infrastructure, incorporating sensing, actuation, control, inspection, SCADA, HMI, and edge computing units over wired TSN links. The section outlines the network architecture, traffic models, node configurations, fault injection methods, and performance metrics used to evaluate system behavior under both normal and failure conditions.

\begin{table}[ht] 
\centering 
\caption{Industrial Network Components}
\resizebox{\linewidth}{!}{ 
\begin{tabular}{|l|p{5.5cm}|} \hline 
\textbf{Component} & \textbf{Function} \\ \hline
PLC (Control Unit) & Central logic controller coordinating sensing, actuation, and robotic tasks. \\ \hline
Sensing Unit & Detects object arrival, triggers inspection, and sends metrics to PLC/Edge unit. \\ \hline
Inspection Camera & Captures high-frequency images and sends them to Edge unit for quality analysis. \\ \hline
Robot Arm & Handles objects, sends status updates to PLC and SCADA. \\ \hline
Actuating Unit & Performs reject/eject operations based on PLC commands. \\ \hline
HMI & Interfaces with operators; receives visual feeds and status information. \\ \hline
SCADA & Supervisory unit for monitoring and control across cells and infrastructure. \\ \hline
Edge unit & Receives and processes data from inspection and sensing units. Sends logs to SCADA. \\ \hline
Cameras (Surveillance) & Broadcast high-rate video streams to SCADA and HMI. \\ \hline
Master Clock & Root clock that distributes time via gPTP across all TSN-enabled devices. \\ \hline
\end{tabular} } 
\label{table:network_components} 
\end{table}

\subsection{Industrial Network Architecture and Configuration}
The industrial network follows a modular, hierarchical architecture designed to emulate realistic industrial automation environments. It comprises two independent production cells, each capable of autonomous operation, and a centralized infrastructure responsible for supervision, monitoring, and edge processing. The network topology includes cell-level TSN switches, central aggregation switches, and interconnecting links configured for redundancy and time synchronization.

Each production cell integrates five key functional units: a sensing unit, actuating unit, robotic arm, inspection system, and a control unit (PLC). These units, detailed in Table \ref{table:network_components}, are dual-homed through two TSN-capable switches (SwitchA\_X and SwitchB\_X, where X denotes the cell index), forming a fault-tolerant local sub-network. This dual-switch design allows for redundant forwarding paths toward central infrastructure via centralSwitch\_1, centralSwitch\_2, and centralSwitch\_3. The backbone consists of 1-Gbps Ethernet links interconnecting switches at the core and aggregation layers. These include:
\begin{itemize}
    \item $centralSwitch\_1$: Core switch connecting both cells.
    \item $centralSwitch\_2$ and $centralSwitch\_3$: Interconnect with SCADA, HMI, Edge unit, and quality-control switches ($qcSwitch\_1$, $qcSwitch\_2$).
\end{itemize}

End devices (e.g., PLCs, sensors, actuators) are connected via 100-Mbps links, representing edge connections commonly found in factory deployments. Figure \ref{fig:network} illustrates this architecture. All switches and hosts are configured with TSN capabilities under the TSNBase configuration. Table~\ref{table:tsn_features} summarizes the TSN features implemented in the network, as configured in the INET-based .ini file.

\begin{table}[ht] 
\centering 
\caption{TSN Features and Their Implementation} 
\resizebox{\linewidth}{!}{ 
\begin{tabular}
{|l|p{5.5cm}|} \hline 
\textbf{TSN Feature} & \textbf{Implementation Details} \\ \hline 
Time Synchronization & IEEE 802.1AS via a dedicated master clock module (masterClock). \\ \hline 
Traffic Scheduling & Always-open gate schedule across all egress ports. \\ \hline 
Credit-Based Shaping & Enabled per-port on all TSN switches. \\ \hline Stream Identification & Per-stream filtering and Priority Code Point (PCP)-based tagging (e.g., 6 for control, 5 for sensing, 2 for cameras). \\ \hline 
Redundancy (FRER) & Frame replication and elimination enabled via StreamRedundancyConfigurator in selected scenarios. \\ \hline 
Per-Stream Filtering & Each TSN switch performs stream classification and rate metering. \\ \hline 
Network Speed & Backbone: 1 Gbps for all inter-switch links; Edge: 100 Mbps for all device links. \\ \hline 
\end{tabular} } 
\label{table:tsn_features} 
\end{table}

This architectural design provides a robust and flexible testbed for evaluating TSN performance in a realistic industrial setting, enabling detailed investigation of latency, jitter, and recovery under fault conditions.

\subsection{Traffic Characterization and Fault Resilience}
The industrial network models a diverse set of traffic streams that reflect the communication requirements of industrial automation systems. These streams are classified by source-destination roles, communication patterns (periodic vs. event-driven), packet size, and criticality level. TSN mechanisms—including stream identification and Priority Code Point (PCP)-based prioritization—are configured to ensure that high-priority traffic receives deterministic service, even under network stress or failure. Table \ref{table:traffic_streams} summarizes key traffic types modeled in the simulation, along with their associated PCP values and timing constraints.

All streams are tagged and encoded using PCP values, as defined by per-stream mapping in the configuration. TSN switches perform ingress classification, policing, and shaping per stream to enforce bandwidth and burst control.

\begin{table*}[t] 
\centering 
\caption{Representative Industrial Traffic Streams}
\resizebox{\linewidth}{!}{
\begin{tabular}{|l|l|c|c|} \hline 
Traffic Type & Source - Destination & PCP & Typical Timing \\ \hline 
PLC Commands & Control\_X - Actuating\_X / RobotArm\_X & 6 & 5-10 ms cyclic \\ \hline 
Sensor Data & Sensing\_X - Control\_X & 6 & 5-10 ms cyclic \\ \hline 
Object Detection & Sensing\_X - Control\_X & 5 & 10–50 ms event-driven \\ \hline 
Inspection Images & Inspection\_X - Edge unit & 5 & 33–100 $\mu s$ periodic \\ \hline 
Barcode/RFID Scans & Inspection\_X - Control\_X & 3 & 10–50 ms event-driven \\ \hline 
Robot Status Feedback & RobotArm\_X - Control\_X & 5 & 5–20 ms periodic \\ \hline 
Camera Surveillance & Camera\_X - SCADA/HMI & 2 & 30 FPS (330 $\mu s$) \\ \hline 
SCADA Reports & Control\_X - SCADA & 4 & 10–100 ms periodic \\ \hline 
Operator Input & HMI - SCADA & 3 & Sporadic / user-driven \\ \hline 
Edge unit Logs & Edge unit - SCADA & 1 & 1–10 sec periodic \\ \hline 
\end{tabular} 
}
\label{table:traffic_streams} 
\end{table*}

To evaluate the fault resilience of the TSN network, two types of failure scenarios are introduced:
\begin{itemize}
    \item S1A1/S1A2 simulate a mid-cell link failure (SwitchA\_1 - SwitchB\_1),
    \item S2A1/S2A2 simulate cell disconnection (centralSwitch\_2 - Edge unit and SCADA).
\end{itemize}

\begin{table*}[ht] 
\centering 
\caption{Fault Injection and Recovery Strategies} 
\resizebox{\linewidth}{!}{ 
\begin{tabular}{|l|p{4.5cm}|p{4.5cm}|} \hline 
\textbf{Fault Type} & \textbf{Description} & \textbf{Recovery Mechanism} \\ \hline 
Link Failure (S1A1/S1A2) & Disconnection between SwitchA\_1 and SwitchB\_1 at 4s. & FRER (S1A2) provides alternate paths using stream replication. \\ \hline 
Cell Disconnection (S2A1/S2A2) & centralSwitch\_2 disconnects from SCADA and Edge unit at 2s. & Redundant inter-switch paths restore communication (S2A2). \\ \hline 
No Redundancy (S1A1/S2A1) & Baseline behavior without FRER or protection. & Traffic interruption occurs with no failover. \\ \hline 
\end{tabular} 
} 
\label{table:fault_injection} 
\end{table*}

Table \ref{table:fault_injection} summarizes these fault scenarios and their recovery logic. The following metrics are used to evaluate network behavior during fault events: \begin{itemize} 
\item \textbf{End-to-End Latency:} Transmission delay from source to destination. 
\item \textbf{Jitter:} Variability in packet delay over time. 
\item \textbf{Packet Loss:} Number of packets dropped during normal and faulted operation. 
\item \textbf{Recovery Time:} Time required to reestablish stream delivery after a failure. \end{itemize}

\begin{algorithm} 
\caption{Initialization of IN2C TSN Simulation} 
\begin{algorithmic}[1] 
\State \textbf{BEGIN Initialization} 
\State Load IN2C topology, TSN configuration, and fault scenario script. \For{each node} 
\State Configure clock with drift model. 
\State Set stream encoder/decoder mappings. 
\State Load UDP application traffic profile (size, interval, PCP). 
\EndFor 
\State Establish topology using TSN switches and assign gPTP roles. 
\State \textbf{END Initialization} 
\end{algorithmic} 
\end{algorithm}

\section{Framework Architecture and Workflow}
The IN2C simulation framework is implemented in OMNeT++ using the INET framework, extended with TSN-specific modules to emulate industrial communication behavior. The simulation models two synchronized production cells and centralized infrastructure, capturing interactions across sensors, PLCs, robot arms, inspection units, SCADA, Edge unit, and HMI systems. INET's TSN extensions—including time-aware switches, gPTP clocks, per-stream filtering, and traffic shaping—enable deterministic and fault-tolerant communication modeling.

All nodes are built from INET’s TsnDevice and StandardHost templates, depending on their role. Each functional unit (e.g., Sensing\_1, Control\_1, Inspection\_2) includes:
\begin{itemize}
    \item A local clock module with oscillator drift,
    \item User Datagram Protocol(UDP) traffic generators and sinks,
    \item Stream coders with PCP-tag mapping,
    \item Bridging and scheduling logic for TSN operation.
\end{itemize}

\noindent Cell switches (SwitchA\_X, SwitchB\_X) and aggregation switches (centralSwitch\_X, qcSwitch\_X) use the TsnSwitch type, with FRER, stream filtering, and shaping enabled as configured.

The master clock (masterClock) operates as a global time reference, distributing synchronization via gPTP across the topology. Each node aligns its local clock to this source, ensuring coherent transmission timing for time-sensitive streams.

The workflow follows an event-driven model:
\begin{itemize}
    \item Initialization: Node clocks are configured, traffic applications instantiated, stream encoders applied, and synchronization established.
    \item Traffic Generation: Application modules produce cyclic or event-driven UDP streams with preconfigured packet size, interval, and destination.
    \item Scheduling: Time-aware and credit-based shaping modules manage queueing and dispatch under bounded latency constraints.
    \item Failure Handling: XML-based ScenarioManager scripts trigger link/node failures (e.g., linkfailure.xml or cellfailure.xml).
    \item Redundancy/Recovery: If redundancy is enabled, FRER-based failover automatically engages. Otherwise, communication may stall.
    \item Metrics Collection: Throughout the simulation, OMNeT++ records delay, jitter, packet drops, and fault recovery durations.
\end{itemize}

Synchronization is driven by IEEE 802.1AS. Each switch forwards sync messages downstream to maintain precise offset and frequency alignment. TSN traffic scheduling uses always-open TAS gates and per-stream credit-based shaping. This configuration is intentionally chosen to isolate the impact of shaping and redundancy mechanisms by avoiding the additional complexity introduced by time-aware gate scheduling. 

\begin{algorithm} 
\caption{Time Synchronization via gPTP} 
\begin{algorithmic}[1] 
\State \textbf{BEGIN TimeSync} 
\For{each TSN node} 
\State Receive gPTP sync from upstream switch. 
\State Adjust local oscillator phase and frequency. 
\State Propagate time to downstream nodes. 
\EndFor 
\State \textbf{END TimeSync}
\end{algorithmic}
\end{algorithm}

Streams are filtered and metered based on source and PCP. Faults are introduced at simulation time using scheduled disconnect/connect events. Redundant configurations trigger alternate paths using FRER trees and packet duplication. This modular, event-driven structure enables controlled experimentation with traffic shaping, synchronization, stream redundancy, and recovery behavior under realistic factory network conditions.

\begin{figure}[t]
    \centering
    \includegraphics[width=\linewidth]{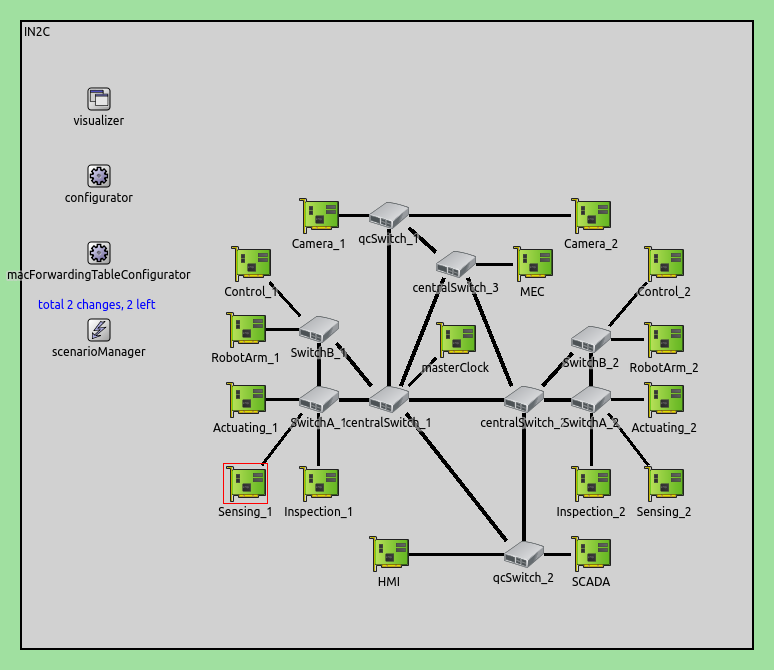}
    \caption{TSN based Industrial Network of Two Production Cells (IN2C).}
    \label{fig:a1net}
\end{figure}

\section{Resilience Evaluation under Network Failures}
This section analyzes the performance of the IN2C TSN network under fault conditions affecting link and cell connectivity. Four scenarios are evaluated to examine the impact of redundancy on latency, jitter, packet loss, and recovery time. Results reflect how TSN mechanisms perform when subject to realistic disruptions in smart manufacturing environments.

\begin{algorithm} 
\caption{TSN Traffic Scheduling and Forwarding} 
\begin{algorithmic}[1] 
\State \textbf{BEGIN Scheduling} 
\For{each output queue} 
\State Filter packets per PCP and stream ID. 
\State Apply shaping (idle slope, burst control). 
\State Dispatch if time gate is open. 
\EndFor 
\State \textbf{END Scheduling} 
\end{algorithmic} 
\end{algorithm}

\begin{algorithm} 
\caption{Fault Injection and Recovery} 
\begin{algorithmic}[1] 
\State \textbf{BEGIN FaultEvent} 
\If{link/node failure triggered} 
\State Disconnect affected modules via gate logic. 
\If{FRER enabled} 
\State Activate alternate stream path. 
\EndIf 
\State Start recovery timer. 
\EndIf 
\If{repair triggered} 
\State Reconnect modules and synchronize clocks. 
\State Log recovery metrics. 
\EndIf 
\State \textbf{END FaultEvent} 
\end{algorithmic} 
\end{algorithm}

\begin{figure}
    \centering
    \includegraphics[width=\linewidth]{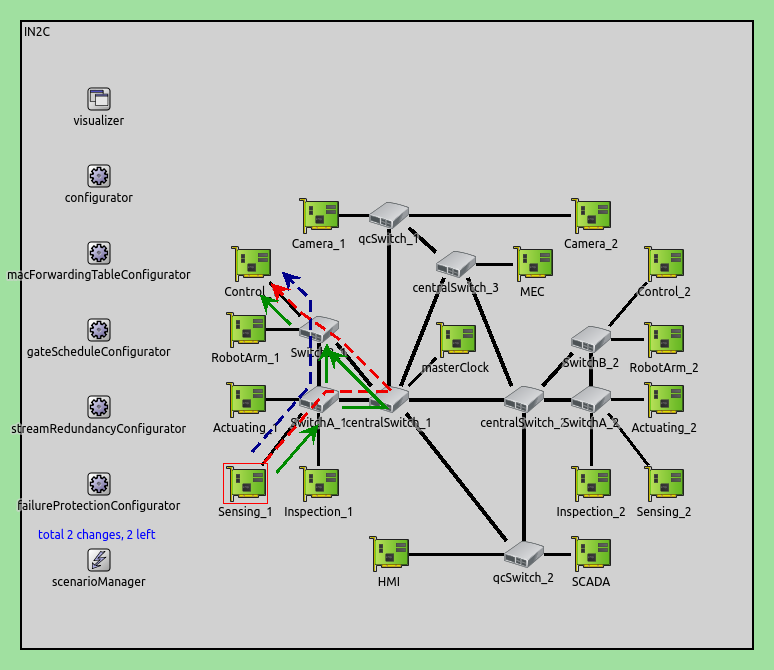}
    \caption{Active FRER to provide link failure protection.}
    \label{fig:s1a2net}
\end{figure}

\begin{figure}
    \centering
    \includegraphics[width=\linewidth]{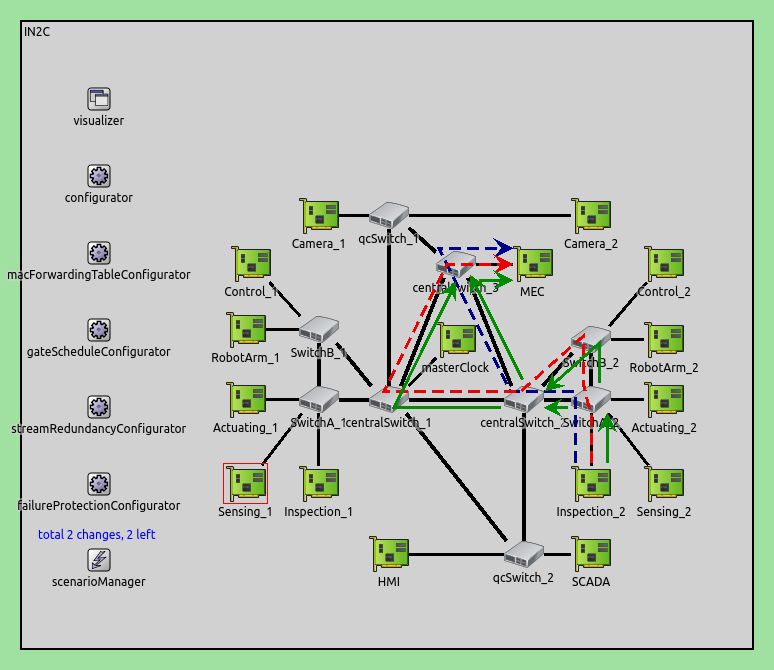}
    \caption{Active FRER to provide cell failure protection.}
    \label{fig:s2a2}
\end{figure}

\subsection{Experimental Setup}
The evaluation is based on the IN2C simulation framework \cite{b12}, developed in OMNeT++ using the INET-TSN extension. The network, Figure \ref{fig:a1net}, consists of two TSN-enabled production cells connected to centralized infrastructure (SCADA, Edge unit, HMI) through a 1-Gbps backbone. End devices communicate via 100-Mbps links.  Each simulation captures behavior across initialization, steady-state, failure, and recovery phases. In this context, four scenarios are defined: 
\begin{itemize} 
\item \textbf{S1A1 / S1A2:} Simulate a link failure between \texttt{SwitchA\_1} and \texttt{SwitchB\_1} at 4 seconds. S1A2, Figure \ref{fig:s1a2net}, enables FRER; S1A1 does not. Runtime: $10 s$.
\item \textbf{S2A1 / S2A2:} Simulate cell isolation by disconnecting \texttt{centralSwitch\_2} from both \texttt{Edge unit} and \texttt{SCADA} at 2 seconds. S2A2, Figure \ref{fig:s2a2}, uses FRER for redundancy. Runtime: $5 s$. \end{itemize}

Failures are introduced using XML-based \texttt{ScenarioManager} scripts, which explicitly disconnect and later restore Ethernet gates between modules. No node-level failures are used; instead, link-level disruptions emulate practical fault cases such as cable damage or aggregation switch malfunction.

Redundancy mechanisms, including Frame Replication and Elimination for Redundancy (IEEE 802.1CB), are activated only in S1A2 and S2A2 via \texttt{StreamRedundancyConfigurator}. Gate control and clock synchronization are implemented using \texttt{EagerGateScheduleConfigurator} and IEEE 802.1AS-compliant gPTP via a centralized \texttt{masterClock}. The full simulation source code, scenarios, and TSN configurations are publicly available at \cite{b12}.

\begin{figure}
    \centering
    \includegraphics[width=\linewidth]{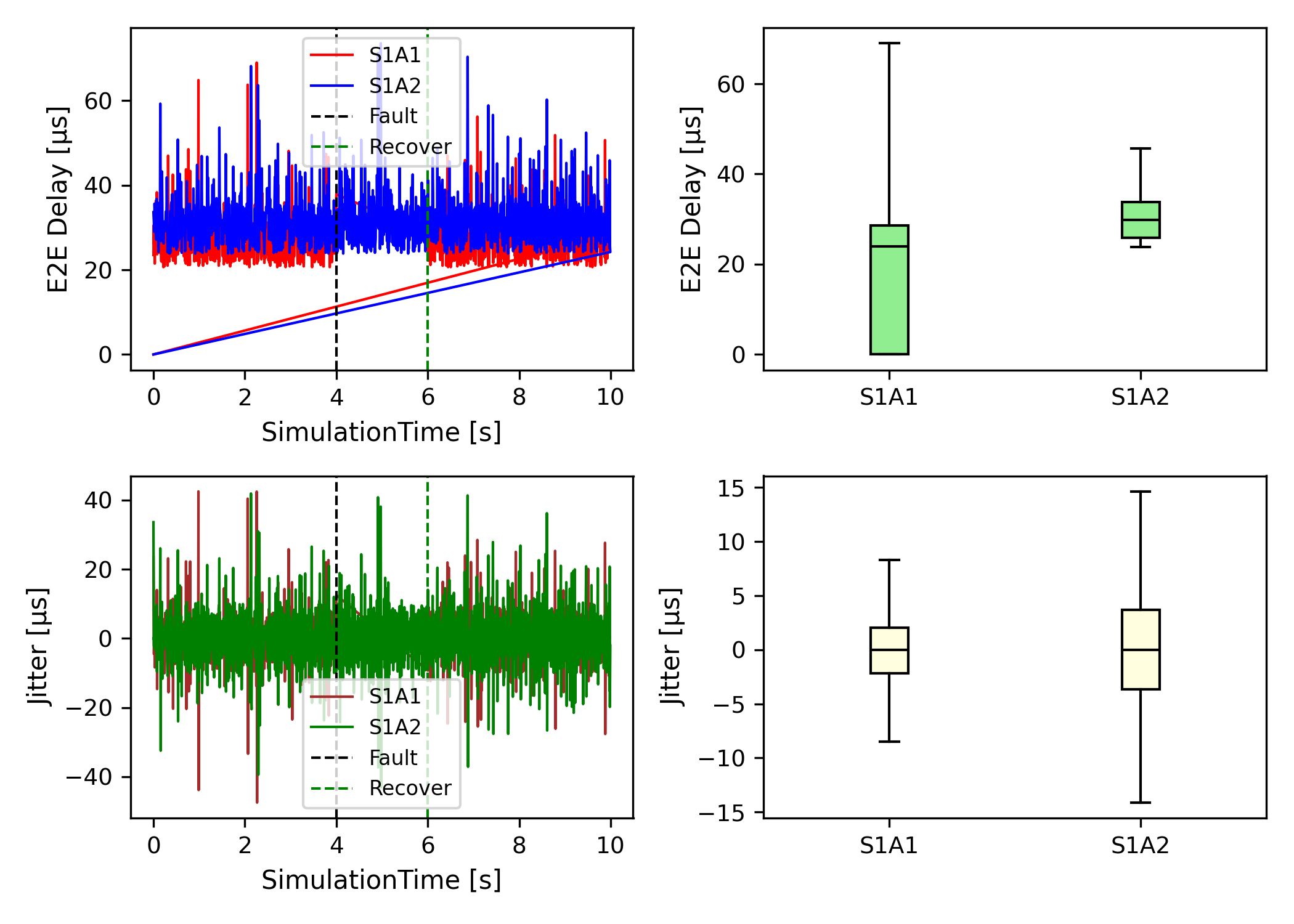}
    \caption{End to End latency and Jitter, Sensing\_1/Control\_1 stream, comparison while handling link failure.}
    \label{fig:s1_latency_jitter}
\end{figure}

\subsection{Latency and Jitter Under Failure}
Figure \ref{fig:s1_latency_jitter} shows the performance of the S1 stream under a link failure between \texttt{SwitchA\_1} and \texttt{SwitchB\_1} injected at $t=4$\ s and resolved at $t=6$\ s. In the non-redundant setup (S1A1), latency becomes highly variable post-failure, with a median of $24.01 \mu$s and a wide spread (mean $18.93 \mu$s). Jitter increases significantly, with a standard deviation of $6.29 \mu$s due to unstable retransmissions and path recovery effects. Despite the existence of physical alternate paths, their absence from the active stream configuration renders the stream vulnerable to the failure.

In contrast, the redundant configuration (S1A2) uses FRER to pre-distribute stream replicas over disjoint paths. This eliminates the need for dynamic rerouting or recovery signaling. As expected,  latency remains consistently bounded despite the fault, though its median rises slightly to $29.81 \mu$s. This increase stems from buffering delay introduced by duplicate frame handling. The trade-off yields a more stable mean delay ($27.43 \mu$s) and enhanced fault tolerance. Jitter remains controlled (std $8.13 \mu$s) as the FRER mechanism ensures that the first valid frame at the receiver is delivered without delay, masking the failure entirely.

\begin{figure}
    \centering
    \includegraphics[width=\linewidth]{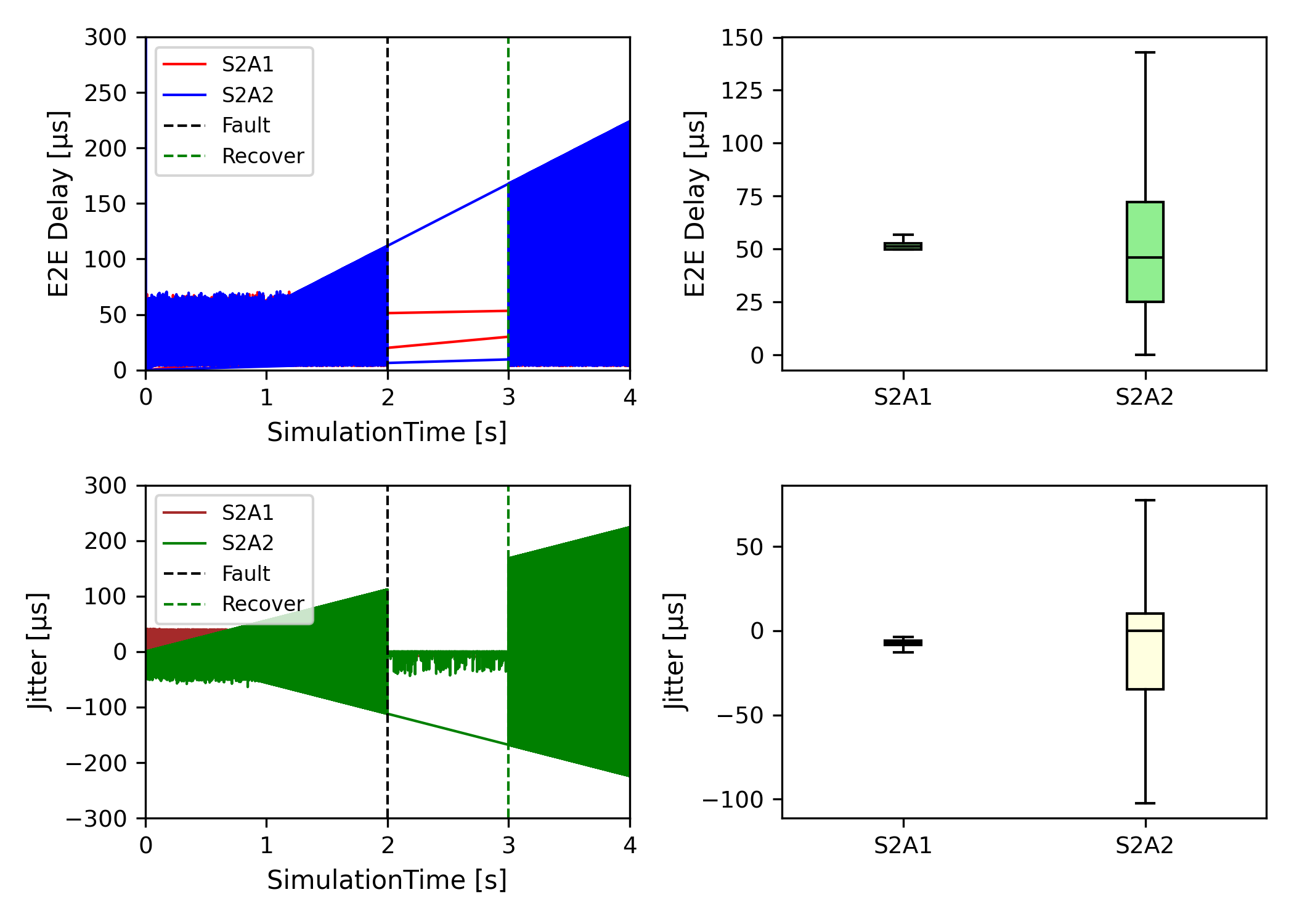}
    \caption{End to End latency and Jitter, Inspection\_2/Edge unit stream, comparison while handling cell failure.}
    \label{fig:s2_latency_jitter}
\end{figure}

Figure \ref{fig:s2_latency_jitter} presents the S2 stream under a cell-level fault, where \texttt{centralSwitch\_2} is disconnected from the Edge and SCADA units at $t=2$\ s and reconnected at $t=3$\ s. 

\begin{table*}[ht]
\centering
\caption{Estimated Packet Loss, Reception, and Loss Rate per Stream}
\resizebox{\linewidth}{!}{ 
\begin{tabular}{|c|c|c|c|c|c|c|c|c|}
\hline
\textbf{Scenario} & \textbf{Stream} & \textbf{Sent} & \textbf{Lost} & \textbf{Received} & \textbf{Loss Rate} & \textbf{Redundancy} & \textbf{Recovery} \\
\hline
S1A1 & Sensing\_1 $\rightarrow$ Control\_1  & 1,337 & 270 & 1,067 & 20.2\% &  No & 2.0 s \\ \hline
S1A2 & Sensing\_1 $\rightarrow$ Control\_1 & 1,331 & 0 & 1,331 & 0.0\% &  FRER & $<$ 1 ms \\\hline
S2A1 & Inspection\_2 $\rightarrow$ Edge unit & 75,096 & 19,776 & 55,320 & 26.3\% &  No & 1.0 s \\\hline
S2A2 & Inspection\_2 $\rightarrow$ Edge unit & 75,126 & 0 & 75,126 & 0.0\% & FRER &  $<$ 1 ms \\
\hline
\end{tabular}
}
\label{tab:loss_rate_analysis}
\end{table*}

In the non-redundant case (S2A1), latency spikes sharply, with a median of $51.28 \mu$s and peaks over 90 $\mu$s. Jitter remains moderate (std $17.32 \mu$s), reflecting consistent recovery behavior once the fault is resolved.

In contrast, S2A2 with FRER exhibits a more complex profile. While the median latency improves slightly to $45.99 \mu$s, the mean increases to $63.27 \mu$s, and the tail becomes pronounced—latency reaches up to $879 \mu$s, and jitter swings from –224 to +224 $\mu$s. This indicates that while FRER masks the failure itself, it introduces significant delivery variability even under normal operation. The congestion between \texttt{centralSwitch\_1} and \texttt{centralSwitch\_3}, likely due to duplicate stream forwarding, amplifies path imbalance and leads to large jitter fluctuations. These results reveal that FRER is not universally beneficial. While it provides failover protection, it may degrade performance for bursty or high-bandwidth streams like inspection traffic.

The right-hand boxplots in both figures confirm these observations statistically. In S1, the non-redundant configuration exhibits outliers and jitter spread during failure, while FRER (S1A2) achieves consistently low variance, validating its effectiveness for critical, low-volume streams. However, in S2, FRER (S2A2) displays a heavy-tailed distribution and wide jitter range, in contrast to the more stable profile of the non-redundant setup (S2A1). These results highlight that while FRER prevents packet loss and ensures continuity, it can introduce delivery unpredictability under normal load, particularly for high-throughput or bursty traffic.

\subsection{Packet Loss and Recovery Analysis}
To quantify the resilience of each configuration, Table~\ref{tab:loss_rate_analysis} summarizes packet transmission, loss, and recovery characteristics for all scenarios.

For the \texttt{Sensing\_1~$\rightarrow$~Control\_1} stream (S1), which transmits every 7.5 ms on average, the S1A1 configuration experiences 270 lost packets during the 2-second link failure—yielding a 20.2\% loss rate. Recovery is purely reactive and delayed until the link is restored. In contrast, FRER-enabled S1A2 maintains continuous delivery across disjoint paths, with no packet loss and instantaneous failover.

\begin{figure}
    \centering
    \includegraphics[width=\linewidth]{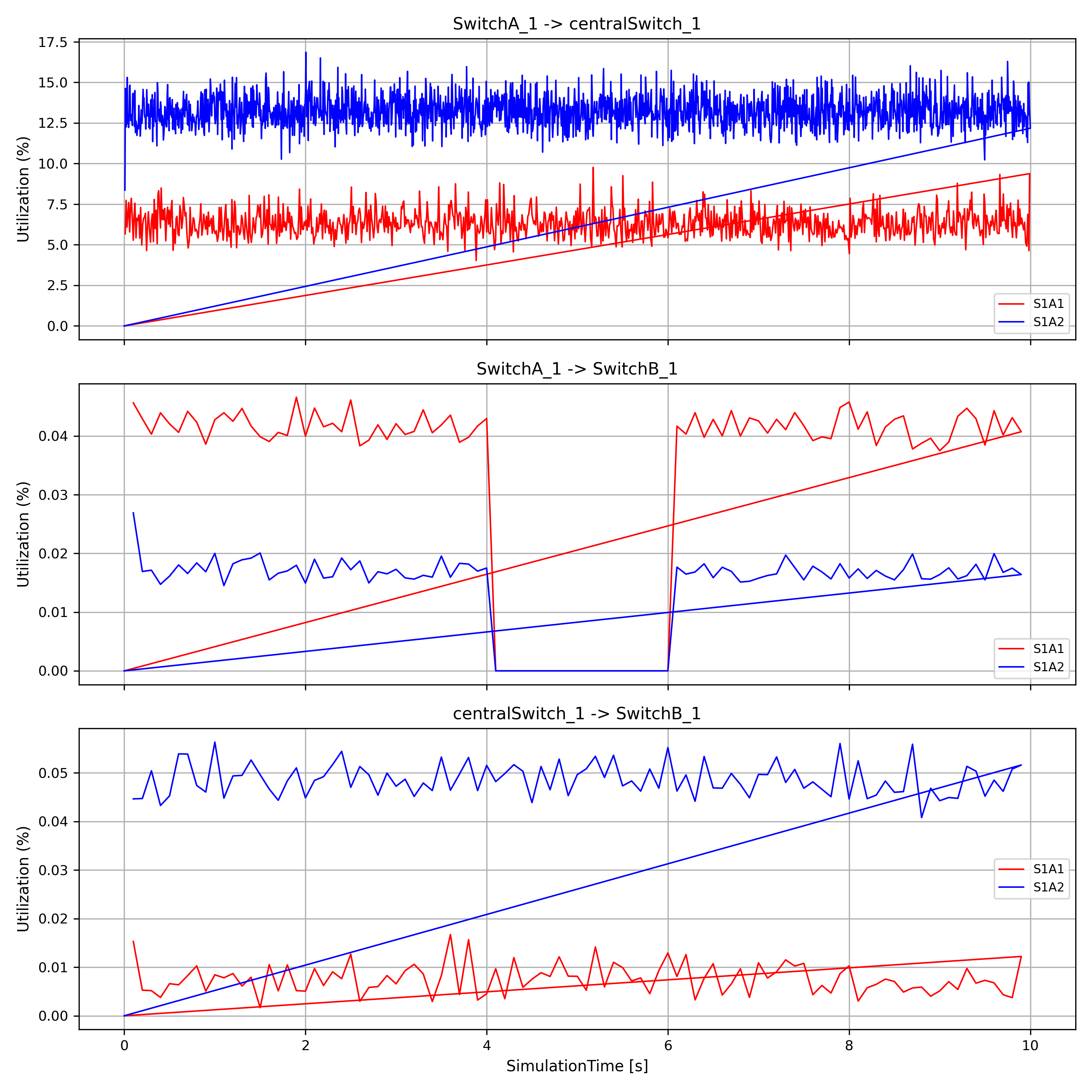}
    \caption{Links utilization for the \texttt{Sensing\_1 $\rightarrow$ Control\_1}. }
    \label{fig:link_utilization_s1}
\end{figure}

\begin{figure}
    \centering
    \includegraphics[width=\linewidth]{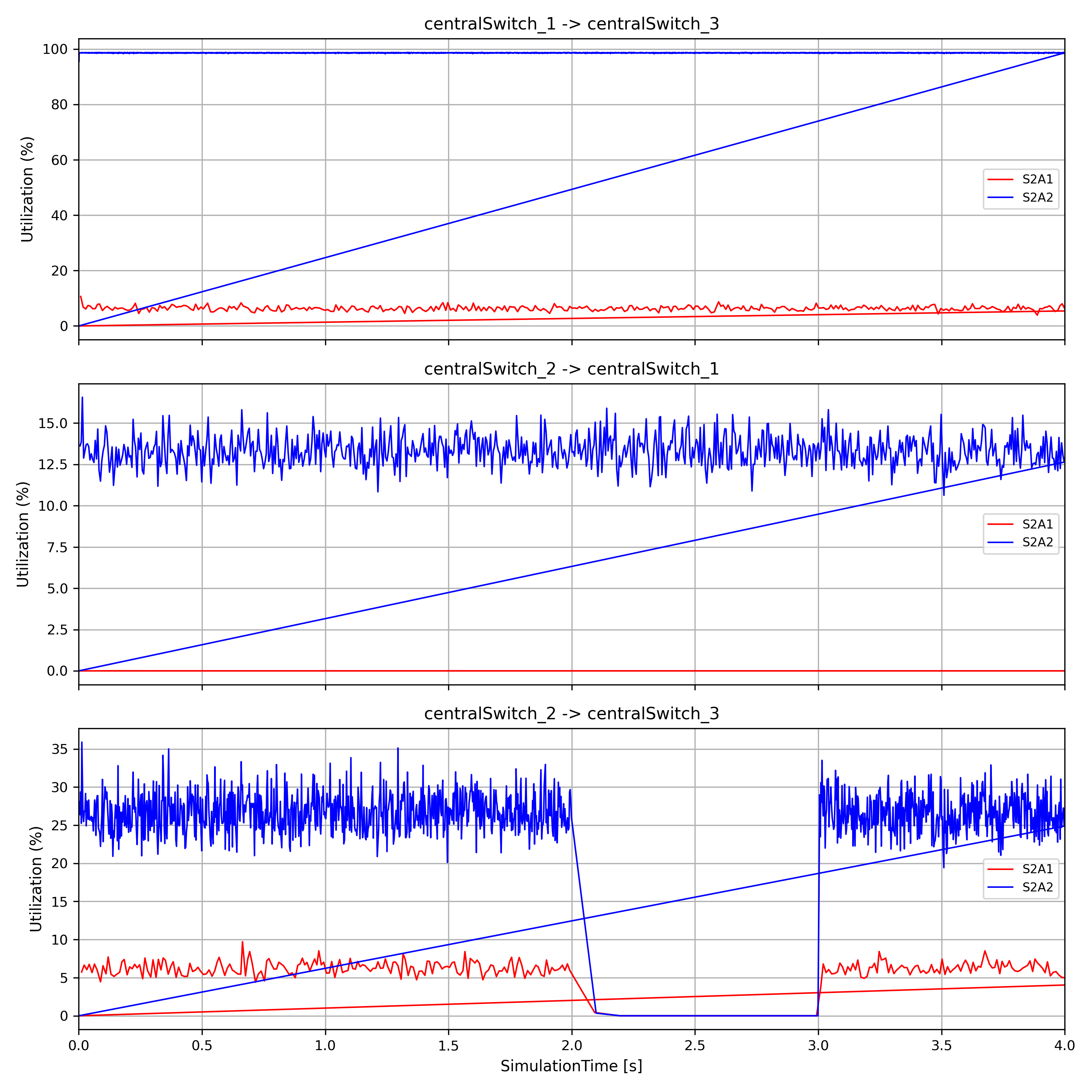}
    \caption{Links utilization for the \texttt{Inspection\_2 $\rightarrow$ Edge unit}.}
    \label{fig:link_utilization_s2}
\end{figure}

The \texttt{Inspection\_2~$\rightarrow$~Edge unit}  stream operates at a high rate (66 $\mu$s intervals), and the S2A1 setup suffers 19,776 dropped packets during a 1-second isolation, corresponding to a 26.3\% loss rate. Despite available physical paths, the absence of redundancy prevents their use. S2A2, leveraging FRER, avoids loss by using duplicate streams; however, as discussed earlier, this comes at the cost of significant jitter and latency variability.

This study compares FRER against a non-redundant baseline, excluding traditional protection schemes like Spanning Tree Protocol (STP), Rapid STP (RSTP), and Ethernet Ring Protection Switching (ERPS), which require tens to hundreds of milliseconds to recover—too slow for Time-Sensitive Networking (TSN) traffic with sub-millisecond deadlines. 
\pagebreak

FRER enables zero recovery time via data-plane replication over disjoint paths but introduces bandwidth overhead and possible jitter due to path imbalance. Alternatives include FRER with Scheduled Traffic (FRER-ST), which improves determinism but needs centralized scheduling; Packet Duplication and Elimination Function (P-DEF) in 5G, which is wireless-specific; and reactive options like Multi-Protocol Label Switching - Transport Profile (MPLS-TP) and IEEE 802.1Qca, which offer fast failover at the cost of added complexity. While FRER is ideal for critical flows, selective or hybrid use is advised for high-bandwidth or bursty traffic.

\subsection{Key Observations}
While FRER eliminates packet loss and enables seamless failover, it introduces measurable bandwidth overhead. Figure \ref{fig:link_utilization_s1} shows link utilization over time for the \texttt{Sensing\_1 $\rightarrow$ SwitchB\_1}, resulting in steady utilization (~4–5\%) until the link failure at $t=4$ s. Utilization then drops to zero, confirming complete disconnection, and only resumes after recovery at $t=6$ s. The lack of alternative forwarding results in data loss.

In S1A2, FRER replicates traffic across disjoint paths—\texttt{SwitchA\_1 $\rightarrow$ centralSwitch\_1} link consistently reaches 13–15\%, while \texttt{centralSwitch\_1$\rightarrow$~SwitchB\_1} carries ~5\%. Meanwhile, the direct path remains active with a reduced load (~2\%). This validates continuous delivery of duplicate frames, confirming the overhead introduced by FRER.

Figure \ref{fig:link_utilization_s2} captures the impact of FRER in the S2 scenario, which involves high-bandwidth streaming from \texttt{Inspection\_2$\rightarrow$ centralSwitch\_3} to near saturation (~100\%). Even during fault-free operation, these links remain heavily loaded due to constant duplication. This congestion not only increases latency and jitter, as shown earlier, but also reduces available headroom for other traffic.

These findings demonstrate that while FRER offers fault resilience, it comes at the cost of substantial link overhead. This trade-off must be carefully considered—especially for high-rate or bursty streams—where redundancy may amplify congestion risks in shared core segments.

\textbf{Recommendation:}
In bandwidth-constrained or converged TSN networks, FRER use must be selective and strategic, prioritizing critical control flows with strict latency and loss constraints. Our evaluation (Section V) shows that indiscriminate replication can increase congestion and jitter, particularly for high-bandwidth streams like inspection video, which experience marginal benefit from FRER unless provisioned with excess capacity. These observations are consistent with prior analyses on redundancy overhead and congestion risk in TSN systems \cite{b10}, \cite{b11}. To mitigate these risks and maintain deterministic performance, we recommend the following:
\begin{itemize}
    \item Topology-aware stream allocation to prevent redundant paths from overlapping at congestion-prone links.
    \item Per-link capacity planning that accounts FRER duplication overhead under both nominal and failure conditions (e.g., up to 2× bandwidth increase).
    \item Flow-aware shaping and policing to bound burst sizes before replication and avoid queue buildup.
    \item VLAN-based isolation to confine traffic classes and avoid cross-stream contention during failover events.
    \item Adaptive redundancy levels, where redundancy is enabled dynamically based on flow criticality, system load, or link health indicators (as explored in \cite{b9}).
\end{itemize}

These measures, informed by both our simulation results and recent studies, help ensure that FRER delivers its intended benefits—lossless failover and zero recovery time—without destabilizing the broader TSN network.

\section{Conclusion and Future Work}
This paper introduced a scenario-driven simulation framework for evaluating TSN resilience under fault conditions in industrial networks. By modeling synchronized production cells with stream-level redundancy, we assessed the impact of link and cell failures on latency, jitter, packet loss, and recovery time.

\pagebreak

Results show that FRER achieves zero-loss failover and uninterrupted operation during faults. However, this reliability comes at the cost of significantly increased bandwidth usage—up to 2–3× in some cases—due to redundant traffic. These findings underscore the need for selective, traffic-aware application of redundancy in bandwidth-constrained environments. 

Future work will enhance the framework with dynamic reconfiguration capabilities, enabling networks to autonomously adapt to faults via gate schedule updates, stream rerouting, and flow prioritization. We also plan to simulate multi-fault and transient failure scenarios to assess TSN’s robustness under more realistic industrial conditions.

\end{document}